\documentclass[reprint,amsmath,amssymb,aps,prl,longbibliography]{revtex4-2}

\usepackage{graphicx}% Include figure files
\usepackage{dcolumn}% Align table columns on decimal point
\usepackage{bm}% bold math
\begin{document}

\preprint{APS/123-QED}
%\title{Statistics of the condensate in large scale quasi-geostrophic turbulence}
%\title{Jet condensation }
\title{Two-dimensional turbulence with local interactions: statistics of the condensate}

%\title{Impact of local interactions on the condensate statistics in two-dimensional turbulence}
\author{Anton Svirsky$^{1}$, Corentin Herbert$^{2}$ and Anna Frishman$^1$}
\email{frishman@technion.ac.il}
% \altaffiliation[ ]{Physics Department, XYZ University.}%Lines break automatically or can be forced with \\
%\email{frishman@technion.ac.il}
%\affiliation{%}%
\affiliation{$^1$Physics Department, Technion Israel Institute of Technology, 32000 Haifa, Israel}
\affiliation{$^2$Univ Lyon, ENS de Lyon, CNRS, Laboratoire de Physique, F-69342 Lyon, France}

\date{\today}% It is always \today, today,

\begin{abstract}
Two-dimensional turbulence self-organizes through a process of energy accumulation at large scales, forming a coherent flow termed a condensate. We study the condensate in a model with local dynamics, the large-scale quasi-geostrophic equation, observed here for the first time. We obtain analytical results for the mean flow and the two-point, second-order correlation functions, and validate them numerically. The condensate state requires parity+time-reversal symmetry breaking. We demonstrate distinct universal mechanisms for the even and odd correlators under this symmetry. We find that the model locality is imprinted in the small scale dynamics, which the condensate spatially confines. 
% Two-dimensional turbulence can self-organize through an out-of-equilibrium process of energy accumulation at large scales. The resulting large, coherent flow is termed a condensate.
% We examine the condensate arising in the large-scale quasi-geostrophic equation, a two-dimensional model with local dynamics never before studied in this regime. We obtain analytical results for the mean flow and the two-point, second-order correlation functions, and validate them against direct numerical simulations. The condensate state requires parity+time-reversal symmetry breaking and our results show distinct mechanisms determining the even and odd correlators under this symmetry. These mechanisms appear to be universal across different two-dimensional flows. At the same time, we find a dramatic influence of the local nature of the dynamics on the condensate state. Regions where the condensate is strong have depleted fluxes towards small scales and a vanishing small-scale dissipation. We explain these observations by the presence of a spatial flux, mediated by the mean flow, which we show to exactly balance injection. 

\end{abstract}
%\keywords{Suggested keywords} %Use showkeys class option if keyword %display desired
\maketitle
%\tableofcontents %\section{\label{sec:Intro} Introduction}
\paragraph{\textbf{Introduction}} Understanding the interactions between turbulent fluctuations and a mean flow is a central problem in fluid mechanics. Such interactions usually prevent a prediction of the mean flow due to their non-linear nature. Recently, some progress was made in this direction for two-dimensional turbulence~\cite{laurie_universal_2014, frishman_turbulence_2018}.
In this case, the mean flow spontaneously emerges from small-scale fluctuations in an out-of-equilibrium process of self-organization. The underlying mechanism is that of an inverse transfer of energy --- from small to large scales, leading to the accumulation of energy at large scales and the establishment of a system-size mean flow termed a condensate~\cite{kraichnan_inertial_1967}. 
%This inverse transfer arises due to the existence of a second inviscid invariant of the system, which is simultaneously transferred to small scales in a so-called direct cascade. 

Two-dimensional flow minimalistically encapsulates the main dynamical constraints imposed by rotation and density stratification, encountered in nature or in laboratory experiments~\cite{xia_spectrally_2009}. 
In fact, a cornerstone of the theory of large-scale dynamics of geophysical flows, including the formation of coherent structures such as jets and vortices, is the quasi-geostrophic (QG) approximation~\cite{vallis_atmospheric_2017}, which shares the same fundamental structure as two-dimensional flow.
However, a general understanding of the feedback between such structures and turbulence within the QG framework is currently lacking. 

In QG, there is a typical scale for the range of fluid elements interactions, called the Rossby radius of deformation $L_d$. The special cases which have been studied in detail so far correspond to dynamics at scales much smaller than $L_d$: the vortex condensate in two-dimensional incompressible Navier-Stokes (2DNS)~\cite{bose_smith_1993,dynamics_chertkov_2007,Chertkov2010,laurie_universal_2014,kolokolov_structure_2016,frishman_turbulence_2018,frishman_culmination_2017}, and jets in the presence of strong differential rotation (beta effect)~\cite{Farrell2007,Srinivasan2012,Tobias2013,Woillez2017, Woillez2019}. Here we address the opposite limit, of scales much larger than $L_d$ (without differential rotation): the large-scale quasi-geostrophic (LQG) equation~\cite{larichev_weakly_1991}. We observe the condensate regime in LQG for the first time in numerical simulations, and provide a detailed theoretical picture of the mean-flow/turbulence interactions. Although the flow is turbulent and the mean flow is sustained due to non-linear interactions, we show that the statistics can be captured within a perturbative approach, the quasi-linear approximation~\cite{marston_recent_2023}. 
This reinforces the relevance of the approach for low-order turbulence statistics in the presence of a strong mean flow~\cite{laurie_universal_2014,frishman_turbulence_2018}. While in 2DNS fluid-element interactions are long-range, in LQG they are local. Comparing the two cases, our work sheds light on the influence of the range of interactions on the condensate state. Indeed, we show that the statistics at large scales are determined  by universal underlying mechanisms. On the other hand, at small scales we find that the locality of interactions has a dramatic influence, with small-scale fluxes and dissipation suppressed in regions where the mean flow is strong.

%A basic understanding of two-dimensional (2D) turbulence is thus fundamental for the understanding of large-scale geophysical and astrophysical flows, e.g.~\cite{rivera_direct_2014, young_forward_2017}. 
% Moreover, turbulence-mean-flow interactions are easier to treat in 2D turbulence as apposed to 3D turbulence, since uncontrolled turbulence-turbulence interactions are not necessary to sustaining the mean flow. Indeed, using a perturbative approach, it was recently possible to obtain the large-scale statistics of the condensate state in 2D incompressible Navier-Stokes (2DNS)~\cite{laurie_universal_2014,frishman_turbulence_2018}. Here we ask how changing the range of interactions between fluid elements, being long range in 2DNS, may influence the condensate state. Are the mechanisms determining the condensate statistics universal? 

% Here we consider the large-scale quasi-geostrophic (LQG) equation --- a 2D flow which supports an inverse cascade, but where interactions between fluid elements are local~\cite{larichev_weakly_1991}. This model can be thought of as an apposing limit to 2DNS. 

\paragraph{\textbf{Framework}} Both the LQG equation and 2DNS can be derived as limiting cases of the shallow water quasi-geostrophic (SWQG) equation. This is an idealized model widely used for flows in the atmosphere and oceans \cite{vallis_atmospheric_2017}, and for magnetically confined plasmas~\cite{diamond_zonal_2005}. In the geophysical context, it captures the dynamics of the free surface of a rapidly rotating shallow fluid layer under the influence of gravity. The inviscid SWQG equation reads
$
\partial_{t}q+\bm{v}\cdot \nabla q=\partial_{t}q+J(\psi,q)=0$, where  $q=\left( \nabla^2 - L_d^{-2}\right)\psi$,
is the potential vorticity, $J(\psi,q)=\partial_{x}\psi\partial_{y}q-\partial_{y}\psi\partial_{x}q$,  and $\psi$ is the stream-function related to the velocity field by $\bm{v} = \bm{\hat{z} \times \bm{\nabla}}\psi$. The flow is assumed to be in geostrophic balance. Thus, $\psi$ is proportional to the deviation of the fluid layer from its mean. 
% This deviation is further assumed to be small in deriving SWQG. 
The system has $L_d$, as a characteristic length scale,
% , given by the speed of a gravity wave divided by the layers' rotation rate,
determining the range of influence of a height perturbation. If the domain size $L\ll L_d$ then every fluid element will influence every other, and the flow will become incompressible, giving the 2DNS. The opposite limit $L_{d}/L\to0$ giving LQG (which can be made consistent with the SWQG approximation~\cite{SI}), corresponds to a very rapidly rotating fluid where surface perturbations remain completely localized. The dynamics become slow in this limit, requiring the re-scaling of time as $\tau=t\left(L_{d}/L\right)^{2}$ (and $\psi$ accordingly), resulting in the LQG equation:
\begin{equation}
\label{eq:LQG}
\partial_\tau\psi + \bm{v}^\omega \cdot \bm{\nabla} \psi=\partial_{\tau}\psi+J(\omega,\psi)=f +\alpha\nabla^{2}\psi-\nu ( -\nabla^{2})^{p}\psi,
\end{equation}
where we defined an effective velocity $\bm{v}^\omega = \bm{\hat{z} \times \bm{\nabla}}\omega$, $f$ is a forcing term, and the last two terms are dissipative --- a viscous term referred to as drag in the following, and a hyper-viscosity. The former provides the dominant dissipation mechanism at large scales, while the latter will dominate at small scales.
The inviscid invariants of \eqref{eq:LQG} are the  kinetic energy $Z=\frac{1}{2}\int\left(\nabla\psi\right)^{2}\text{d}^{2}x=\frac{1}{2}\int\bm{v}^{2}\text{d}^{2}x$ and all moments of $\psi$, in particular the potential energy
$E=\frac{1}{2}\int\psi^{2}\text{d}^{2}x$. The existence of two positive quadratic invariants results in the inverse cascade of $E$ and a direct cascade (from large to small scales) of $Z$~\cite{smith_turbulent_2002}. 

The LQG advection equation is similar to 2DNS but with the roles of the vorticity $\omega = \nabla^2 \psi$ and the stream-function reversed.
% --- the vorticity acting as the "effective stream-function" for $\bm{v}^\omega $. 
Indeed, LQG and 2DNS are part of a class of active scalar equations where a scalar $q$ is advected by a velocity with stream-function $\phi$, and $q_{\bm{k}} = |{\bm{k}}|^m \phi_{\bm{k}}$~\cite{pierrehumbert_spectra_1994}, $m$ controlling the range of the dynamics~\cite{venaille_violent_2015}. For 2DNS, $m=2$ and the dynamics is long range. For LQG $m=-2$, so that the advecting velocity is determined by active scalar gradients, making the dynamics local. 
% Another geophysically relevant case is surface quasigeostrophic dynamics~\cite{held_surface_1995} which has $m=1$. Fluid particle interactions are then long range but are more local than in 2DNS~\cite{venaille_violent_2015}.

% To get a statistically steady state condensate we add forcing and dissipation terms to \eqref{eq:LQG free}:
% \begin{equation}
% \label{eq:LQG}
%     \partial_{\tau}\psi+J( \omega,\psi ) = f +\alpha\nabla^{2}\psi-\nu ( -\nabla^{2})^{p}\psi
% \end{equation}
% where $f$ is a random forcing acting in a narrow band of scales $l_f$, $\alpha$ is the large scale drag (linear drag on velocity) halting the inverse cascade of potential energy and $\nu$ is the small scale (hyper) viscosity halting the kinetic energy cascade. 

\paragraph{\textbf{Simulations}} 
Direct Numerical Simulations (DNS) are performed by integrating \eqref{eq:LQG} using the Dedalus framework \cite{burns_dedalus_2020}. 
% The pseudo-spectral method is implemented using the 3/2 dealiasing rule and time stepping using a third-order, four-stage DIRK/ERK method.
Simulating in a doubly periodic box of dimensions $L\equiv L_{y}=2L_x=2\pi$, we enforce the formation of a jet-type condensate by breaking the symmetry between the $x$ and $y$ directions~\cite{bouchet_random_2009,frishman_jets_2017}.
%We focus on a jet-type LQG condensate, simplifying the following analysis. In a doubly periodic domain such a condensate emerges if the symmetry between the $x$ and $y$ directions is broken~\cite{bouchet_random_2009,frishman_jets_2017}. We therefore use a box of dimensions $L\equiv L_{y}=2L_x=2\pi$, with spatial resolution of $64\times128$, restricted by the rapid decrease of the eddy-turnover time with decreasing scale in LQG~\cite{SI}. 
We use a spatial resolution of $64\times128$, a white-in-time forcing with characteristic length scale $l_f = 2\pi/13$
%which is localized in Fourier space at a wavenumber $k_{f} =2\pi/l_f= 13$. 
% (forcing in an annulus of width $2dk=2$ with a constant amplitude $A=10^{-3}$ and a random phase). 
% Results from simulations with a larger/smaller forcing scale are presented in~\cite{SI}. 
, hyper-viscosity with $p=7$ and $\nu =7.3\times10^{-19}$ and take $\alpha=(0.5,1,2)\times 10^{-3}$. 
%Parameters are chosen such that a significant fraction of the (potential) energy is transferred to large scales, requiring $Re\equiv l_f^{2p-8/3} \epsilon^{1/3}/\nu \gg 1$, where $\epsilon=\langle f \psi\rangle$ is the energy injection rate. 
Condensation of $E$ at large scales requires a slow dissipation rate compared to the eddy-turnover time at the box scale, amounting to the condition $\delta \equiv \alpha (L^{2}/\epsilon)^{1/3} \ll1$, where $\epsilon=\langle f \psi\rangle$ is the energy injection rate.  Additional simulation details, including for DNS with other parameters, are in~\cite{SI}.  
%The full list of simulations performed is presented in Table~\ref{tab:sim-list}. 

\begin{figure}[b t!]
\includegraphics[width=1\columnwidth]{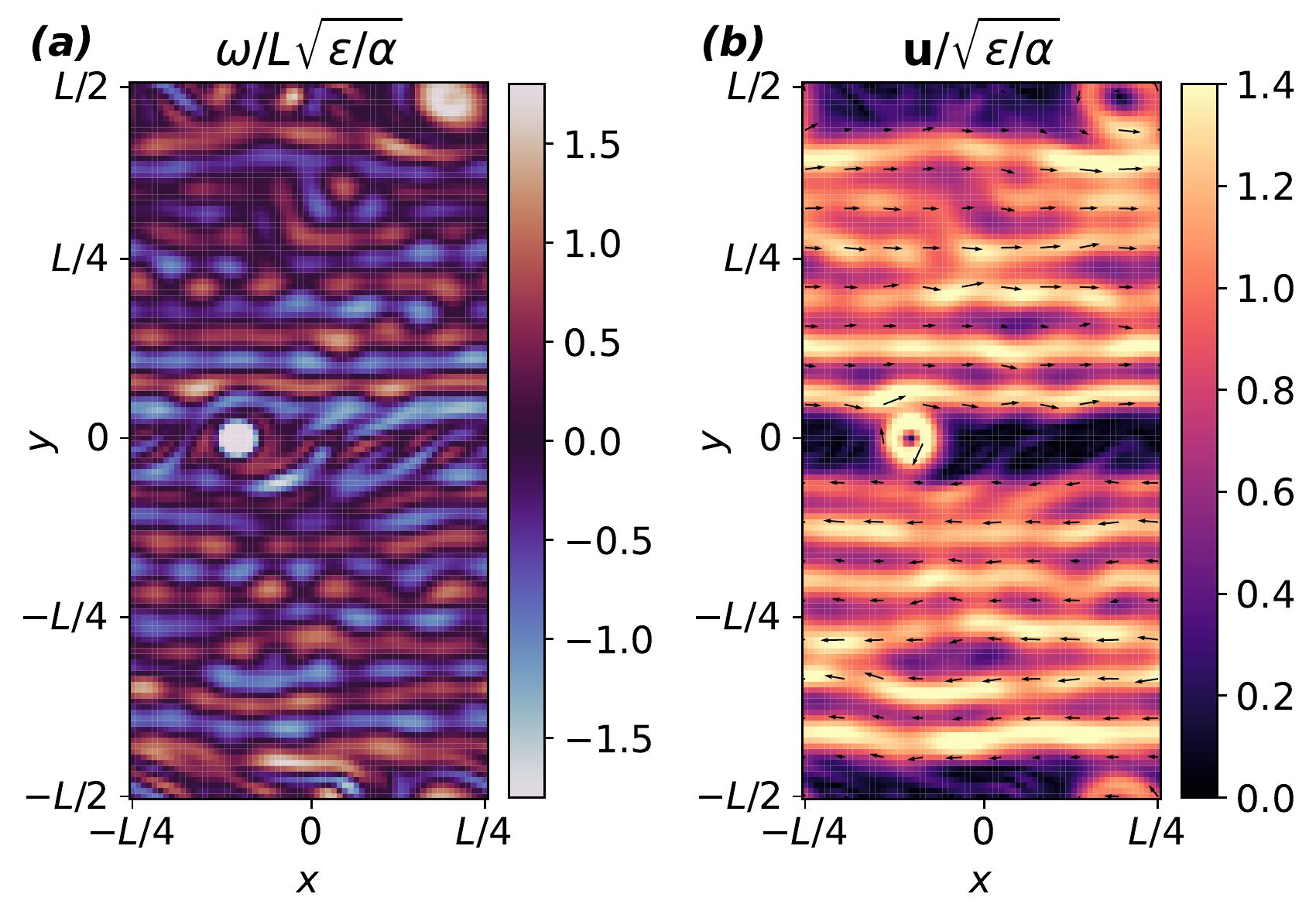}
\caption{\label{fig:Snapshot} LQG jet condensate in a periodic, rectangular box with $\delta = 0.052$. (a) Vorticity snapshot $\omega=\nabla^{2}\psi$ and (b) Velocity snapshot $\boldsymbol{v}=\hat{z}\times\boldsymbol{\nabla}\psi$.}
\end{figure}

The steady state condensate takes the form of two alternating jets along the short side ($x$ direction) of the domain, as demonstrated in Fig.~\ref{fig:Snapshot}. Here the jet structure is selected by the domain geometry, unlike in beta-plane turbulence. Between the jets, there are two small vortices, similarly to what was found in 2DNS~\cite{frishman_jets_2017}. 
% In the jet region, the flow is statistically homogeneous in $x$. 
% Small magnitude oscillations along the $y$ direction of the jet amplitude can also be observed in Fig.~\ref{fig:Snapshot}.  

%\section{Leading order sol. for the mean flow}
\paragraph{\textbf{Mean flow}} We now set out to obtain a statistical description of the steady state LQG jet condensate. We start by decomposing the flow into the mean $\Psi = \langle \psi \rangle$ and fluctuations $\psi^\prime = \psi - \Psi$. We will focus on the jet region, where the flow is statistically homogeneous in $x$, and the mean flow depends on $y$ only. In particular, the mean flow is in the $x$ direction: $U=-\partial_y \Psi$. In the steady state, averaging \eqref{eq:LQG} and neglecting the influence of hyper-viscosity at large scales we obtain
\begin{equation}
\label{eq:ReAvg LQG}
\partial_{y}\left[
\left\langle v_{y}^{\omega\prime}\psi^{\prime}\right\rangle  - \alpha\partial_y \Psi\right]=0,
\end{equation}
where we have used that $V_{y}^{\omega}  = \partial_x \langle \omega \rangle = 0$. As mentioned before, $\psi$ is proportional to the layer height, thus \eqref{eq:ReAvg LQG} can be interpreted as a mass balance and $\langle v_y^{\omega\prime} \psi^\prime \rangle$ as the turbulent mass flux, analogously to the momentum flux or Reynolds stress in 2DNS. We note that a constant non-zero total flux is inconsistent with the symmetries of our system, as there is no preferred direction aside from the one imposed by the emerging mean flow, giving
\begin{equation}
\label{eq:RS}
\left\langle v_{y}^{\omega\prime}\psi^{\prime}\right\rangle =\alpha\partial_{y}\Psi.
\end{equation}
Next, we obtain the potential energy balance of the mean flow by multiplying \eqref{eq:LQG} by $\Psi$ and averaging:
\begin{equation}
\label{eq:ReAvg E MF}
\partial_{y}\left[\Psi\left\langle v_{y}^{\omega\prime}\psi^{\prime}\right\rangle 
- \alpha \Psi \partial_y \Psi \right] =\left\langle v_{y}^{\omega\prime}\psi^{\prime}\right\rangle \partial_{y}\Psi-\alpha\left(\partial_y \Psi\right)^{2}.
\end{equation}
Similarly, we obtain the potential energy balance of fluctuations by multiplying \eqref{eq:LQG} by $\psi^\prime$ and averaging, giving:
\begin{equation}
\label{eq:ReAvg E flac}
\partial_{y}\left[
\langle v_{y}^{\omega\prime}\frac{\psi^{\prime 2}}{2}\rangle - \alpha \langle \psi^\prime \partial_y \psi^\prime \rangle\right]
= \epsilon - \langle v_{y}^{\omega\prime} \psi^{\prime}\rangle\partial_{y}\Psi.
\end{equation}
In writing \eqref{eq:ReAvg E flac} we have neglected the dissipation of potential energy by fluctuations, since the development of the condensate implies dissipation by fluctuations is inefficient and thus small. Equations \eqref{eq:ReAvg E MF} and \eqref{eq:ReAvg E flac} reflect the flow of potential energy in the system, where the interaction term $\langle v_{y}^{\omega\prime} \psi^{\prime}\rangle \partial_{y}\Psi=-U \langle v_{y}^{\omega\prime} \psi^{\prime}\rangle$ controls the out-of-equilibrium transfer of energy between the fluctuations and the mean flow. The inverse nature of potential energy transfer in this system implies that it goes \textit{from} the fluctuations \textit{to} the mean-flow. This requires mass-flux up the gradient of the mean height: $\langle\psi^{\prime}\bm{v}^{\omega\prime}\rangle\cdot\bm{\nabla}\Psi>0$. 

Due to relation \eqref{eq:RS} the left hand side of equation \eqref{eq:ReAvg E MF}, corresponding to the spatial flux of mean potential energy $\Psi^2$, vanishes. Thus, here unlike in 2DNS~\cite{laurie_universal_2014}, all the energy input from the fluctuations is dissipated locally by the mean flow. Note that the transfer of potential energy to the mean flow cannot be the dominant process in regions where $\partial_{y}\Psi=-U\approx0$ since $\langle v_{y}^{\omega\prime}\psi^{\prime}\rangle\partial_{y}\Psi$ is small there. For jets in a periodic domain there always must be such a region as $\int\bm{\nabla}\Psi\text{d}^{2}x=0$ identically.  

Finally, to close the system of equations we assume that interactions with the mean flow are much faster than nonlinear interactions between fluctuations, $\left| \langle v_{y}^{\omega\prime}\psi^{\prime}\rangle\partial_{y}\Psi \right| \gg \left| \langle\psi^{\prime} \bm{v}^{\omega\prime}\cdot\bm{\nabla}\psi^{\prime}\rangle \right|$, justified by the development of a strong condensate for $\delta\ll1$.
This is called the quasi-linear approximation, which has a long history~\cite{marston_recent_2023}, and was recently successfully used in a similar context for 2DNS~\cite{laurie_universal_2014}. Unlike 2DNS, in LQG the dynamics are local (e.g. no pressure term), and thus so is the analysis, obviating the need for further approximations.  

\begin{figure}[b t!]
\includegraphics[width=1\columnwidth]{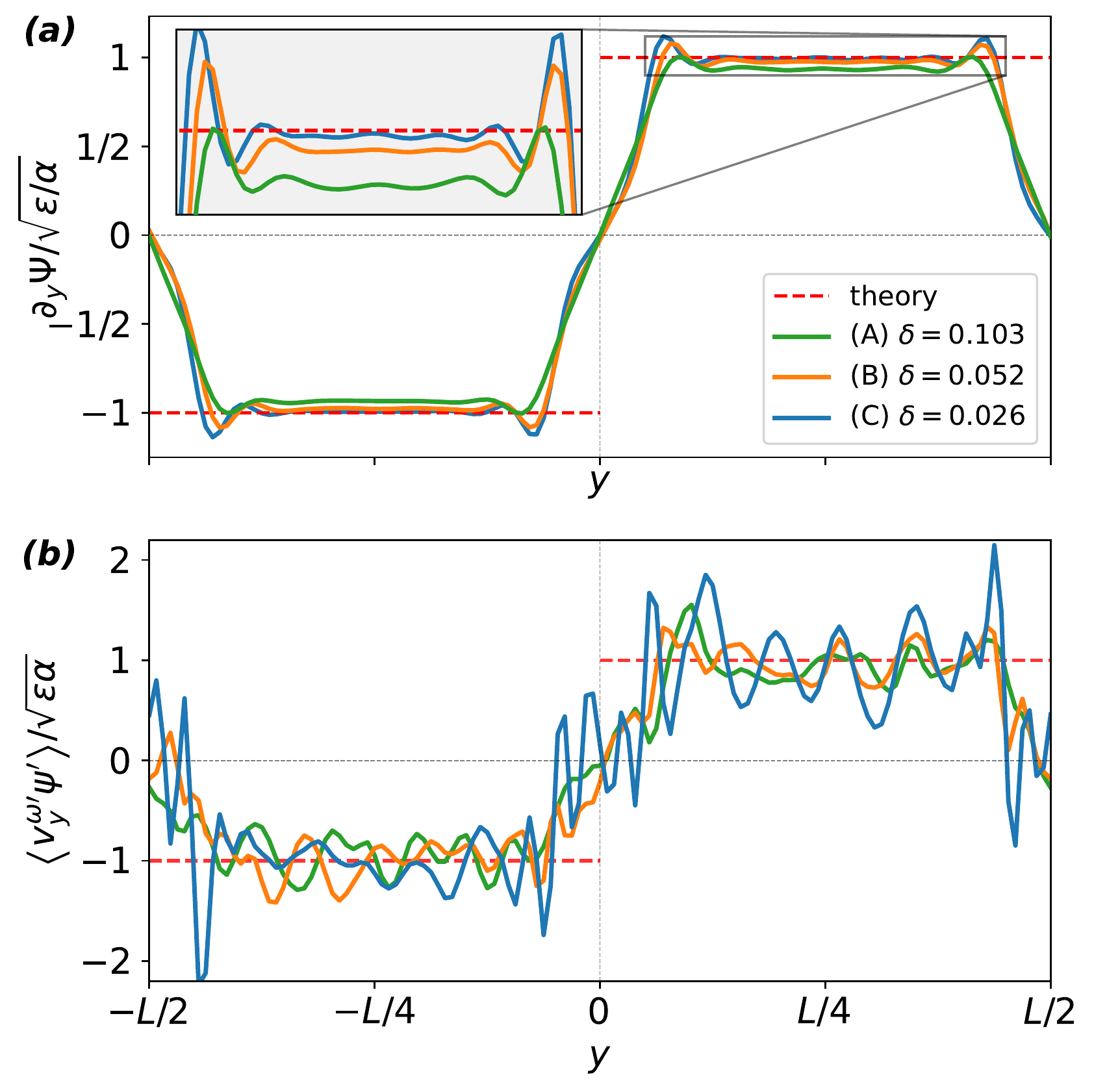}
\caption{\label{fig:U+RS} (a) Horizontal velocity profile $\partial_y\Psi$, dashed line is the prediction \eqref{eq:JetSol U}; (b) Mass flux $\langle v_y^{\omega\prime} \psi^\prime \rangle$,  dashed line is the prediction \eqref{eq:JetSol pv}.}
\end{figure}

Using \eqref{eq:RS} together with  \eqref{eq:ReAvg E flac} in the quasi-linear approximation gives the leading order solution
\begin{eqnarray}
\partial_{y}\Psi =-U= \pm\sqrt{\frac{\epsilon}{\alpha}}, \label{eq:JetSol U}
\\
\label{eq:JetSol pv}
\langle\psi^{\prime}v_{y}^{\omega\prime}\rangle	= \pm\sqrt{\epsilon\alpha},
\end{eqnarray}
implying a flat mean velocity profile $U$ (and mass flux) for each of the jets. % This also implies that the mean effective velocity $V^{\omega}_x=0$ to leading order. 
 We compare these predictions to results from DNS in Fig.~\ref{fig:U+RS}. For the averaging procedure, after reaching steady state, we perform a temporal (over $\sim 1000$ large scale turnover-time) as well as a spatial average along the $x$-axis, utilizing the  homogeneity of the system. For the mean velocity profile, good agreement with \eqref{eq:JetSol U} can be seen in Fig.~\ref{fig:U+RS}(a), demonstrating convergence to the leading order prediction with decreasing $\delta$.
The main difference between DNS and the theoretical prediction is the oscillations of the jet velocity seen in DNS (see also Fig. \ref{fig:Snapshot}), which persist after averaging over long times. They appear to be a sub-leading correction to the mean flow profile $U$ (controlled by a small parameter other than $\delta$), not captured at leading order. The mass flux $\langle\psi^{\prime}v_{y}^{\omega\prime}\rangle$ is also computed from DNS, shown in Fig.~\ref{fig:U+RS}(b). It fluctuates about the theoretical prediction, due to rapid fluctuations of $v_{y}^{\omega\prime}$ requiring additional statistics for convergence. Still, the sign of the mass flux always matches the sign of $\partial_y\Psi$ (in the jet region), as is necessary for the inverse transfer. 

\paragraph{\textbf{Two-point correlation function}} Having obtained the mean flow profile, we now examine the two-point (single-time) correlation function $\left\langle \psi_{1}^{\prime}\psi_{2}^{\prime}\right\rangle \equiv\left\langle \psi^{\prime}(\bm{r_1})\psi^{\prime}(\bm{r_2})\right\rangle$ where $\bm{r_i}=(x_{i},y_{i})$. We obtain an equation for $\left\langle \psi_{1}^{\prime}\psi_{2}^{\prime}\right\rangle$ at leading order in perturbation theory from \eqref{eq:LQG}, neglecting dissipation of the fluctuations, using the statistical homogeneity in $x$, and the quasi-linear approximation to neglect higher order moments~\cite{SI}. In addition, we use that from Eq.~\eqref{eq:JetSol U}, $V^{\omega}_x=0$ at leading order. We thus get
\begin{equation}
\begin{split}
\label{eq:2-pt equation full}
    \left[\mathcal{L}_{1}+\mathcal{L}_{2}\right]\left\langle \psi_{1}^{\prime}\psi_{2}^{\prime}\right\rangle &=2\epsilon\Phi_{12},
    \end{split}
\end{equation}
where the operator $\mathcal{L}_i=\partial_{y_i}\Psi(y_i)\nabla^{2}\partial_{y_i}$ (no summation is implied here) and $\left\langle f(\bm{r_1},t)f(\bm{r_2},t')\right\rangle =2\epsilon\Phi_{12} \delta(t-t')$ is the force two-point correlation function with $\Phi_{12}\equiv\Phi(\Delta x/l_{f},\Delta y/l_{f})$. Using the mean profile Eq.~\ref{eq:JetSol U}, this equation reads
\begin{equation}
\label{eq:2-pt equation simple}
    \partial_{y_{+}}\partial_{y_{-}}\partial_{x_{1}}\left\langle \psi_{1}^{\prime}\psi_{2}^{\prime}\right\rangle =2\sqrt{\epsilon\alpha}\Phi_{12},
\end{equation}
where we have defined the variables $y_{+}=\frac{y_{1}+y_{2}}{2}$ and $y_{-} =\Delta{y}/2=\frac{y_{1}-y_{2}}{2}$. Integrating \eqref{eq:2-pt equation simple} we get:
\begin{equation}
    \label{eq:2p-solution}
    \left\langle \psi_{1}^{\prime}\psi_{2}^{\prime}\right\rangle =C(\Delta x,\Delta y)+2y_{+}\sqrt{\alpha\epsilon}l_f^2\int_{0}^{\frac{\Delta x}{l_{f}}}dz\int_{0}^{\frac{\Delta y}{2l_{f}}}dz'\tilde{\Phi}\left(z,z'\right),
\end{equation}
with $\tilde{\Phi}(x,y) = \Phi(x,y)-\hat{\Phi}(k_x=0,y)-\hat{\Phi}(x,k_y=0)$, subtracting the contribution to the forcing from $k_x=0$ and $k_y=0$ modes~\cite{SI}. The second term in~\eqref{eq:2p-solution} is the inhomogeneous solution to equation~\eqref{eq:2-pt equation simple}, due to a balance between the forcing and the mean flow. 
The first term, $C(\Delta x,\Delta y)$, is a homogeneous solution of~\eqref{eq:2p-solution}, i.e. a zero mode of the advection operator $\mathcal{L}_{1}+\mathcal{L}_{2} = \partial_{y_{+}}\partial_{y_{-}}\partial_{x_{1}}$. This family of zero modes is chosen assuming decaying correlations with $\Delta y\to L/2$, and using the result from DNS that the variance $\langle(\psi')^{2}\rangle$ is constant in the jet region~\cite{SI}. Deriving the functional form and amplitude of the zero modes is beyond the scope of this work. 
Empirically, our DNS results~\cite{SI} point to the scaling $\langle u^{\prime 2}\rangle = -\partial_{\Delta y}^2 C|_{(0,0)} \sim(\epsilon L)^{2/3} \delta^{-1/2}$ and $\langle v^{\prime 2}\rangle = -\partial_{\Delta x}^2 C|_{(0,0)} \sim(\epsilon L)^{2/3} \delta^{1/4}$, while from Eq.~\eqref{eq:JetSol pv} the off-diagonal term $\langle u^\prime v^\prime \rangle = (\epsilon L)^{2/3} (y/L) \delta^{1/2}$. Thus, the perturbation theory is indeed consistent, at most $\langle u^{\prime 2}\rangle/U^2 \sim \delta^{1/2} \ll1$. 
%Our DNS results also point to the scaling $\langle(\psi')^{2}\rangle=C(0,0)\sim(\epsilon L^{4})^{2/3}$ (for fixed viscosity and $l_f$)~\cite{SI}, i.e that the variance is independent of $\delta$. Thus, the perturbation theory is indeed consistent, $\langle(\psi')^{2}\rangle/U^2L^2\sim \delta \ll1$. 
% For our parameters, the inhomogeneous solution is further suppressed compared to the zero modes, their ratio scaling as $\delta^{1/2}(l_f^2/L^2)(l_f^2/l_\nu^2)\ll1$. 
 
% \begin{figure}[bt]
% \includegraphics[width=1\columnwidth]{3_var+2p.pdf}
% \caption{\label{fig:zmodes+var} Fluctuations statistics: (a) The variance, dashed lines mark the jet boundary. (b) The correlation function $\langle \psi_1^\prime \psi_2^\prime \rangle$ taken at $y_+=0$ and normalized by its variance, demonstrating the symmetry with respect to $\Delta x\to -\Delta x$. (simulation B, averaged over both jets)}
% \end{figure}
%Having obtained the two-point function, we are able to compute all the odd (under parity) single-point quantities of the form $\left\langle a^{\prime}b^{\prime}\right\rangle$  under the assumption they originate from the inhomogeneous term. Specifically the value of $\langle v_{y}^{\omega\prime}\psi^{\prime}\rangle$ and other single point quantities agrees with \eqref{eq:JetSol pv} obtained previusly.\\

\paragraph{\textbf{PT symmetry breaking}} The condensate state owes its existence to out-of-equilibrium fluxes which break time-reversal symmetry. On the level of single time correlators such symmetry breaking will manifest itself through a combined parity+time reversal (PT), $x\to-x$ and $t\to-t$, breaking. Indeed, the inviscid LQG dynamics with a homogeneous in $x$ mean flow is statistically invariant under PT. However, the mass flux $\langle\psi^{\prime}v_{y}^{\omega\prime}\rangle$ (responsible for the inverse transfer of energy) is odd under this symmetry and can be non-zero only if it is broken. 

We propose that even and odd under PT correlators are determined through different mechanisms, reflected in the decomposition of the two-point correlator~\eqref{eq:2p-solution} into, respectively, the zero modes and inhomogeneous solution. The inhomogeneous solution is determined by a balance between the forcing, which breaks time reversal symmetry, and the mean flow, which breaks parity (but preserves PT). The homogeneous solution, on the other hand, is insensitive to the forcing and instead is a zero mode of the advection operator $\mathcal{L}_1+\mathcal{L}_2$. A similar structure was inferred in 2DNS~\cite{kolokolov_structure_2016,frishman_culmination_2017,frishman_turbulence_2018}. 

The inhomogeneous term in~\eqref{eq:2p-solution} is indeed odd under PT: it changes sign under reflection $x\to-x$ (so that $\Delta x\to -\Delta x$) and, being a single-time quantity, does not explicitly depend on time. We propose that it fully determines odd correlators. The mass flux $\langle\psi^{\prime}v_{y}^{\omega\prime}\rangle$ can indeed be directly computed from it, in agreement with~\eqref{eq:JetSol pv}~\cite{SI}. The limit of large scale-separation $L/l_f\gg1$ is interesting to consider at this point, as in this limit we expect universal behaviour, independent of forcing details.
This is manifested in Eq.~\eqref{eq:2p-solution}: the influence of the inhomogeneous term becomes limited to separations $\Delta x, \Delta y\lesssim l_f$, meaning that its main role becomes to determine single-point (odd) correlation functions~\cite{SI}. 

% This is a consequence of $\tilde{\Phi}_{12}$ appearing in~\eqref{eq:2p-solution}, instead of $\Phi_{12}$, giving a vanishing contribution for larger separations~\cite{SI}. 

Turning to even-under-PT correlators, if they are determined by zero modes then those should be even under PT, implying $C(-\Delta x,\Delta y)=C(\Delta x,\Delta y)$.
% (note that $C(-\Delta x,-\Delta y)=C(\Delta x,\Delta y)$ trivially due to the symmetry of the two-point correlator to the exchange $1\leftrightarrow 2$). 
We verify this property, as well as that the even part of $\left\langle \psi_{1}^{\prime}\psi_{2}^{\prime}\right\rangle$ is independent of $y_+$ in the jet region, as expected for the zero modes, in~\cite{SI}. 
% Note that in our DNS the even contribution to $\left\langle \psi_{1}^{\prime}\psi_{2}^{\prime}\right\rangle$ everywhere dominates the odd part~\cite{SI}.   

\begin{figure}[bt!]
\includegraphics[width=1\columnwidth]{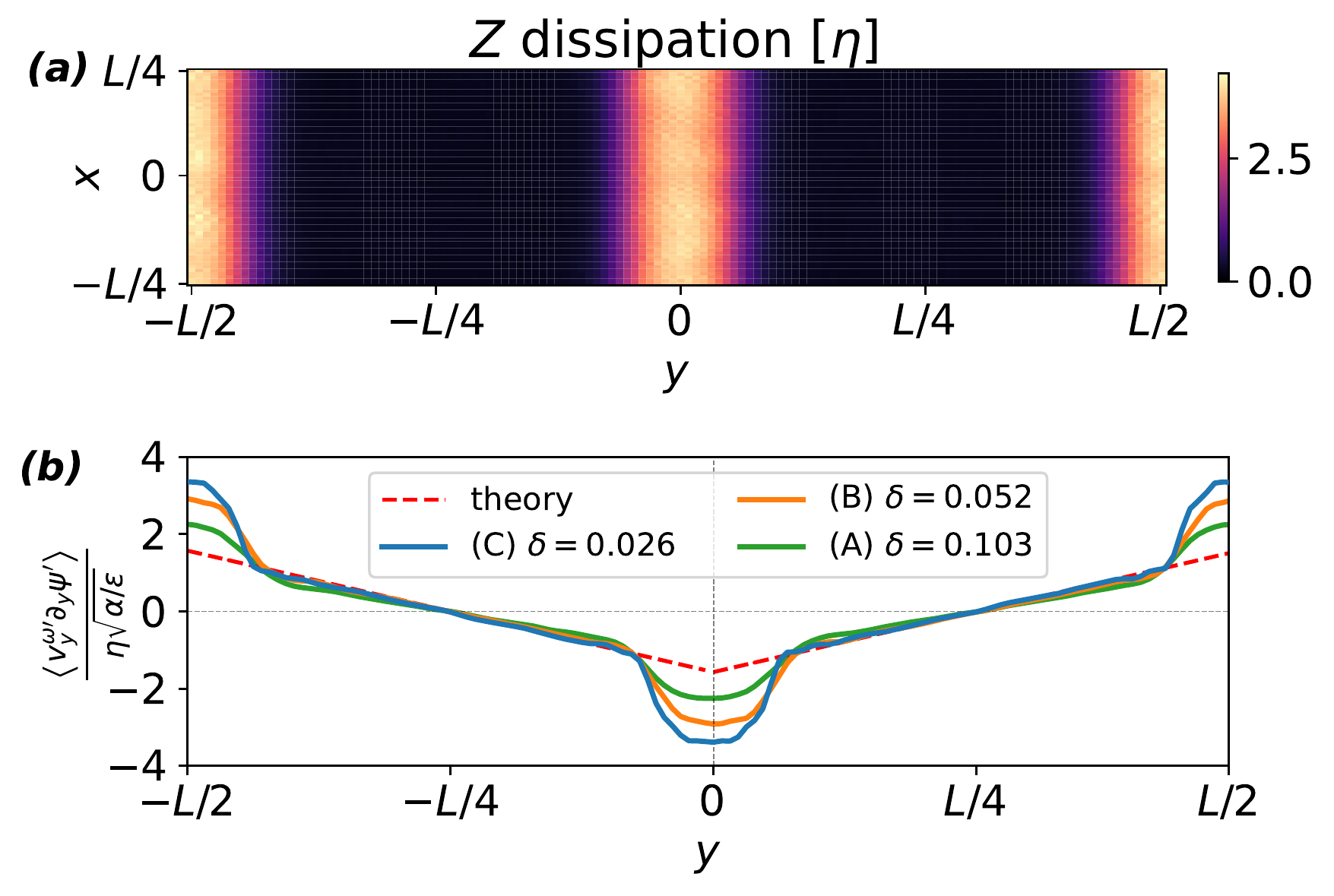}
\caption{\label{fig:diss+ke_flux} (a) Mean kinetic energy dissipation normalized by $\eta$; (b) The term $\left\langle v_{y}^{\omega\prime}v_{x}^{\prime}\right\rangle$ as measured from DNS, the theoretical prediction \eqref{eq:k-en flux} is marked by the dashed line.}
\end{figure}

\paragraph{\textbf{Arrest of the direct cascade}} The condensate necessarily coexists with a direct cascade to small scales of a second inviscidly conserved quantity---the kinetic energy in LQG. As the direct cascade involves scales from the forcing scale and smaller, we generally expect the mean flow to have a negligible effect on it. This is indeed the case in 2DNS~\cite{frishman_jets_2017} where the direct cascade of enstrophy (squared vorticity) remains homogeneous and isotropic, as reflected in an essentially spatially uniform enstrophy dissipation rate. Surprisingly, we find this is not the case in LQG, as seen in Fig.~\ref{fig:diss+ke_flux}(a): the kinetic energy dissipation is concentrated in regions in between the jets, being close to zero elsewhere. To explain this inhomogeneity we turn to the kinetic energy balance, obtained by multiplying the derivative $\partial_{i}$ of \eqref{eq:LQG} by $\partial_{i}\psi^{\prime}$ and averaging. In the steady state, the balance of kinetic energy for the fluctuations reads~\cite{SI}: 
\begin{equation}
    \label{eq:KE balance}
   \partial_{y}\left[J_{y}'+ I_D'\right] = \eta - D'+T,
\end{equation}
where $\eta$ is the kinetic energy injection rate, $I_D'$ and $D'$ originate from the viscous terms and are the respective flux and dissipation rate, $T=\partial_yU \partial_y\langle v_{y}^{\omega\prime} \psi' \rangle $ is a transfer term between the mean flow and the fluctuations, and $J_y'$ is a spatial kinetic energy flux given by 
\begin{equation}
\label{eq:Jy}
J_{y}' =U\langle v_{y}^{\omega\prime} u^\prime \rangle
   +
   \langle\partial_{y}\psi^{\prime}\bm{v}^{\omega\prime} \cdot \bm{\nabla}\psi^{\prime}\rangle-\langle\omega^{\prime}v_{y}^{\omega\prime}\psi^{\prime}\rangle.
\end{equation}
%Using that $\langle \omega^\prime v^{\omega\prime}_y\rangle = \partial_x \langle (\omega^\prime)^2\rangle /2 = 0$ we get that 
 The transfer term $T$ both here and in 2DNS can be estimated to generically be much smaller than the injection rate $\eta$~\cite{SI}. It identically vanishes for the leading order solution~\eqref{eq:JetSol U},\eqref{eq:JetSol pv}. Furthermore, we expect the kinetic energy fluctuations to have homogeneous statistics at small scales (where they reside), so that spatial fluxes involving solely fluctuations should be small. We thus expect the balance
 \begin{equation}
     \partial_{y}\left[U\langle v_{y}^{\omega\prime} u \rangle\right] = \eta - D'.
     \label{eq:KE_f}
 \end{equation}
 The key difference between Eq.~\eqref{eq:KE_f} and the analogous balance in 2DNS is in the spatial flux due to the mean flow, preventing a local balance between injection and dissipation. It is this spatial flux which arrests the direct cascade in regions where the mean flow $U$ is strong, carrying the injected kinetic energy away from those regions. By a direct computation using \eqref{eq:2p-solution}, see~\cite{SI}, we find that 
\begin{equation}
    \label{eq:k-en flux}
    U\langle v_{y}^{\omega\prime} u^\prime \rangle=\partial_y 
    \Psi \langle\partial_{y}\psi^{\prime}v_{y}^{\omega\prime}\rangle 
    =
    \eta y.
\end{equation}
%where we have used $\eta = \epsilon k_f^2$.
This is confirmed by a direct comparison to DNS in the jet region, Fig.~\ref{fig:diss+ke_flux}(b).
Combining~\eqref{eq:KE balance} and~\eqref{eq:k-en flux} we find that the spatial flux carries all the injected kinetic energy before it has time to cascade to small scales and dissipate there, in agreement with Fig.~\ref{fig:diss+ke_flux}(a).  

\paragraph{\textbf{Conclusion}} We presented results from a perturbative analysis and direct simulations of LQG in the condensate regime, observed here for the first time. Obtaining the two-point correlation functions, we have revealed the different mechanisms determining the odd and even correlators under parity+time reversal. While the former are determined at small scales by a balance between the forcing (breaking time reversal) and the mean flow (breaking parity), the latter are constrained to be zero modes of an advection operator. In fact, this is also the underlying structure in 2DNS~\cite{kolokolov_structure_2016,frishman_culmination_2017,frishman_turbulence_2018}, suggesting a universal picture across different 2D flows.
We also find an unexpected influence of the mean flow on the direct cascade, which is locally arrested in regions with a strong inverse energy transfer. This phenomenon, absent in 2DNS, is explained by the formation of an out-of-equilibrium spatial flux out of regions where the mean flow is strong. This flux term is a consequence of the locality of the dynamics, and is absent for active scalar models with $m>0$ and for SWQG with $L_d>l_f$~\cite{SI}. It is however not ruled out for SWQG with $L_d<l_f$, of which LQG is a limiting case, and it will be interesting to see if a similar expulsion of the direct cascade could be observed there. 

\textbf{Acknowledgments:} A.F. would like to thank Guido Boffettta and Stefano Musacchio, and acknowledge the 2017 SCGP workshop on fluid flows, for nucleating the idea for this project.
% \begin{figure}[h]
% \includegraphics[width=1\columnwidth]{Figures/diss.pdf}
% \caption{\label{fig:Jet_cond_disipation} Mean kinetic (a) and potential (b) energy dissipation normalized by $\eta$ and $\epsilon$ respectively.}
% \end{figure}

% \begin{figure}[ht]
% %\includegraphics[width=0.9\columnwidth]{Figures/joint-B.png}
% \includegraphics[width=0.95\columnwidth]{Figures/c2+kflux_v2.pdf}
% \caption{\label{fig:k-en flux} (a) Demonstration of flatness of the fluctuations potential energy in the jet region, as predicted by \eqref{eq:2p-solution}; (b) The term $\left\langle v_{y}^{\omega\prime}v_{x}^{\prime}\right\rangle$ as measured from the DNS against the theoretical prediction given by \eqref{eq:k-en flux}}
% \end{figure}
%\section{Conclusion}

%\bibliographystyle{apsrev4-2}
%

%\nocite{*}
\end{document}